\documentclass{mem}
\usepackage{natbib}\usepackage{txfonts}\usepackage{balance}
\usepackage{graphicx}
\usepackage[a4paper,breaklinks,dvipdfm]{hyperref}
\idline{75}{282}

\usepackage{color}
\begin{document}
\def\teff{$T\rm_{eff }$}
\def\kms{$\mathrm {km s}^{-1}$}

\title{
Long-term spectral study of the black hole Cygnus X-1 using INTEGRAL
}

   \subtitle{}

\author{
F. Cangemi\inst{1} \and T. Beuchert\inst{2} \and T. Siegert \inst{3} \and V. Grinberg \inst{4} \and J. Wilms \inst{5} \and J. Rodriguez \inst{1} \and \\ I. Kreykenbohm \inst{5} \and P. Laurent  \inst{6} \and K. Pottschmidt \inst{7, 8}
          }
\institute{
\footnotesize
CEA/AIM Paris Saclay, 91191 Gif sur Yvette, France
\and
API/UVA, Science Park 904, 1098 XH Amsterdam, The Netherlands
\and
CASS/UCSD, 9500 Gilman Drive, La Jolla, CA 92093, USA
\and 
 IAAT, Universit\"at Tübingen, Sand 1, 72076 T\"ubingen, Germany
\and
Dr. Remeis Sternwarte \& ECAP, FAU, Sternwartstr. 7, 96049 Bamberg, Germany 
\and
Laboratoire APC, UMR 7164, CEA/CNRS/Université Paris Diderot, 75013 Paris, France
\and
NASA Goddard Space Flight Center,Greenbelt, MD 20771, USA
\and
University of Maryland, Baltimore County, Baltimore, MD 21250, USA
}

\authorrunning{F. Cangemi, T. Beuchert}

\titlerunning{Cygnus X-1 using INTEGRAL}

\abstract{We utilize the joint capabilities of IBIS and SPI to perform a state-resolved 20--2000\,keV analysis of the microquasar Cygnus X-1. In both LHS and HSS, the spectral analysis reveals the presence of a high-energy tail above 400\,keV in addition to a standard, Compton-like continuum in the 20--400\,keV range. We study the polarisation properties of the hard X-ray radiation, as well as of the polarisation of this high-energy tail. Polarisation is detected in the LHS in agreement with previous work. We find potential variations of the LHS polarisation with time and no polarisation detection in the HSS.

}
\maketitle{}

\begin{figure*}[t!]
\resizebox{\hsize}{!}{\includegraphics[clip=true]{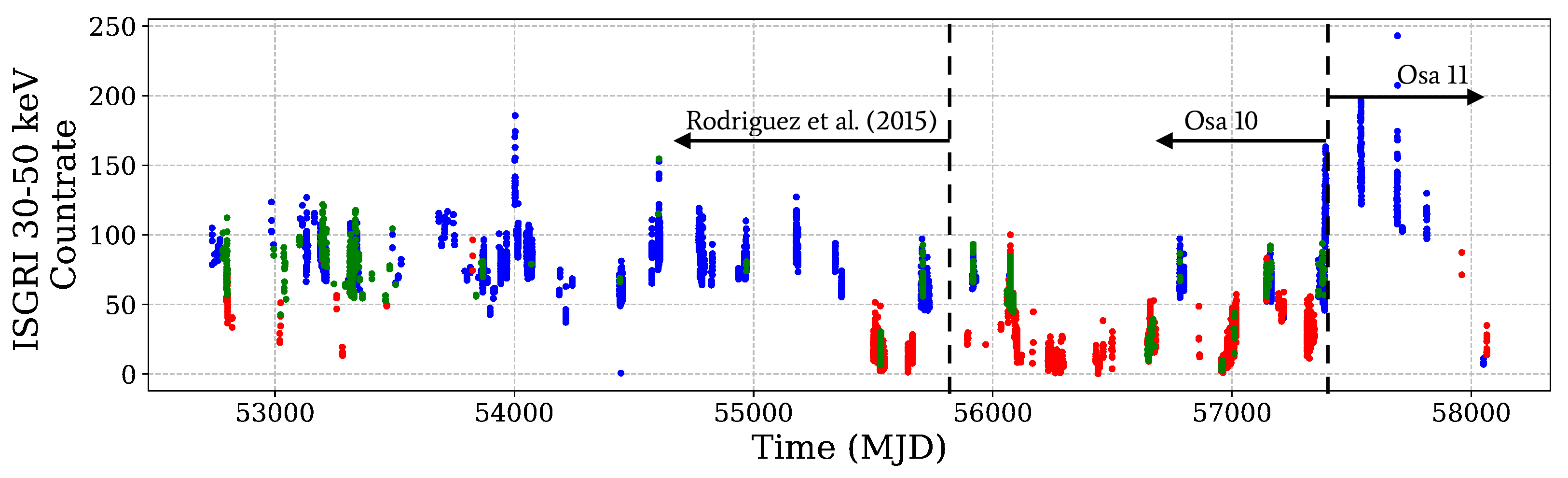}}
\vspace{-0.5cm}
\caption{\footnotesize 
ISGRI ScW-wise 30--50 keV light curve. The LHS, HSS and IS ScW are represented in blue, red and green respectively. The arrows indicate the \texttt{OSA} change and data used previously.}
\label{lc}
\end{figure*}

\section{Introduction}
X-ray binaries show two main spectral states: the ``Low Hard State", LHS, dominated by hard X-rays and the ``High Soft State", HSS, dominated by soft X-rays. The LHS spectrum can be described by a hard power law (photon index $\Gamma<2$) with an exponential cutoff around 100 keV, which indicates a dominance of inverse Compton scattering of a hot plasma off a lower-energy photon field. The HSS spectrum is well modelled by a steeper power law ($\Gamma>$2.5) together with a black body like spectrum which describes an optically thick accretion disc close to the black hole. In the radio domain, jets have been observed in the LHS \citep[e.g.][]{corbel_formation_2013}, but they seem absent or quenched in the HSS.
 The transition from the LHS to the HSS is accompanied with discrete ejections at relativistic velocities \citep[e.g.][]{mirabel_microquasars_1998, rodriguez_2_2008}. 

Beyond this, a high-energy tail from 400\,keV up to several MeV has been detected in multiple sources \citep{grove_gamma-ray_1998}. In Cyg~X-1, it was first seen by CGRO/COMPTEL \citep{McConnell2000}, later confirmed with INTEGRAL \citep[hereafter L11, R15 for the last two]{bel_broad-band_2006, jourdain_separation_2012, jourdain_emission_2012, laurent_polarized_2011, rodriguez_spectral_2015}. In HSS, further observations are needed to confirm the tail. The origin of the tail is debated; possible interpretations are synchrotron emission from the jets, or a hybrid thermal/non-thermal corona.

Here, we extend earlier work presented in L11 and R15. We gathered 15 years of data with INTEGRAL and took advantage of the source changing into HSS around 2012 and remaining in that state for considerable time. 
We exploit the joint capabilities of IBIS and SPI to constraint the high-energy tail in the LHS and, for the first time, collect enough data to investigate the spectrum $>$400 keV in the HSS. Using the Compton mode, we can use polarisation as an independant diagnostic to characterise the origin of the high energy component.

\section{Observations and long-term behaviour}

We consider all INTEGRAL data from MJD\,52722--58064. The details of the data reduction will be given in Cangemi et al. (in prep). Since the \texttt{OSA} software has been updated to its \texttt{11.0} version in October 2018, which is only valid for data taken after MJD 57388, the ISGRI extraction has to be divided in two parts, see Fig.~\ref{lc}. We used the method of \cite{grinberg_long_2013} to classify each Science Window (ScW) into LHS, intermediate state (IS) and HSS. 

Fig.\,\ref{lc} shows the long-term ISGRI light curve.
Cyg X-1 mostly remained in LHS until MJD 55700 and transited on MJD 55786 into a $\sim$4 years long HSS. 
It transited back in the LHS around MJD 57531.

We stacked individual ScW depending on their assigned state and the \texttt{OSA} version used due to inconsistent calibration between \texttt{OSA 10.2} and \texttt{OSA 11.0}. 

\section{Phenomenological spectral analysis}

The ISGRI spectra were modelled in the 25--400 keV (\texttt{OSA~10.2}) and 30--400 keV (\texttt{OSA~11.0}) range and the SPI spectra in the 25--2000 keV ranges.
 
\textbf{LHS}. We first restricted our analysis to energies below 400 keV. With \texttt{const*cutoffpl}, we obtained an adequate fit (113.5/89 $\chi^2$/dof) with a photon index of $\Gamma_{\mathrm{\mathrm{cut}}}$=1.50$\pm$0.02 and an exponential cutoff of $E_{\mathrm{cut}}$=157$\pm$5 keV. An extra reflection component is statistically not required. 
We then added the data above 400\,keV and froze $\Gamma_{\mathrm{\mathrm{cut}}}$. We do confirm hard X-ray excess emission as previsouly reported by L11, R15. This high-energy tail component is well described with a power law with $\Gamma_{\mathrm{po}}$=1.5$^{+0.3}_{-0.5}$. Once again, no reflection is needed (147.7/98 for \texttt{const*reflect(cutoffpl+po)} versus 147.83/97 $\chi^2$/dof for \texttt{const*(cutoffpl+po)}).

\begin{figure}[]
\resizebox{.5\textwidth}{!}{\includegraphics[clip=true]{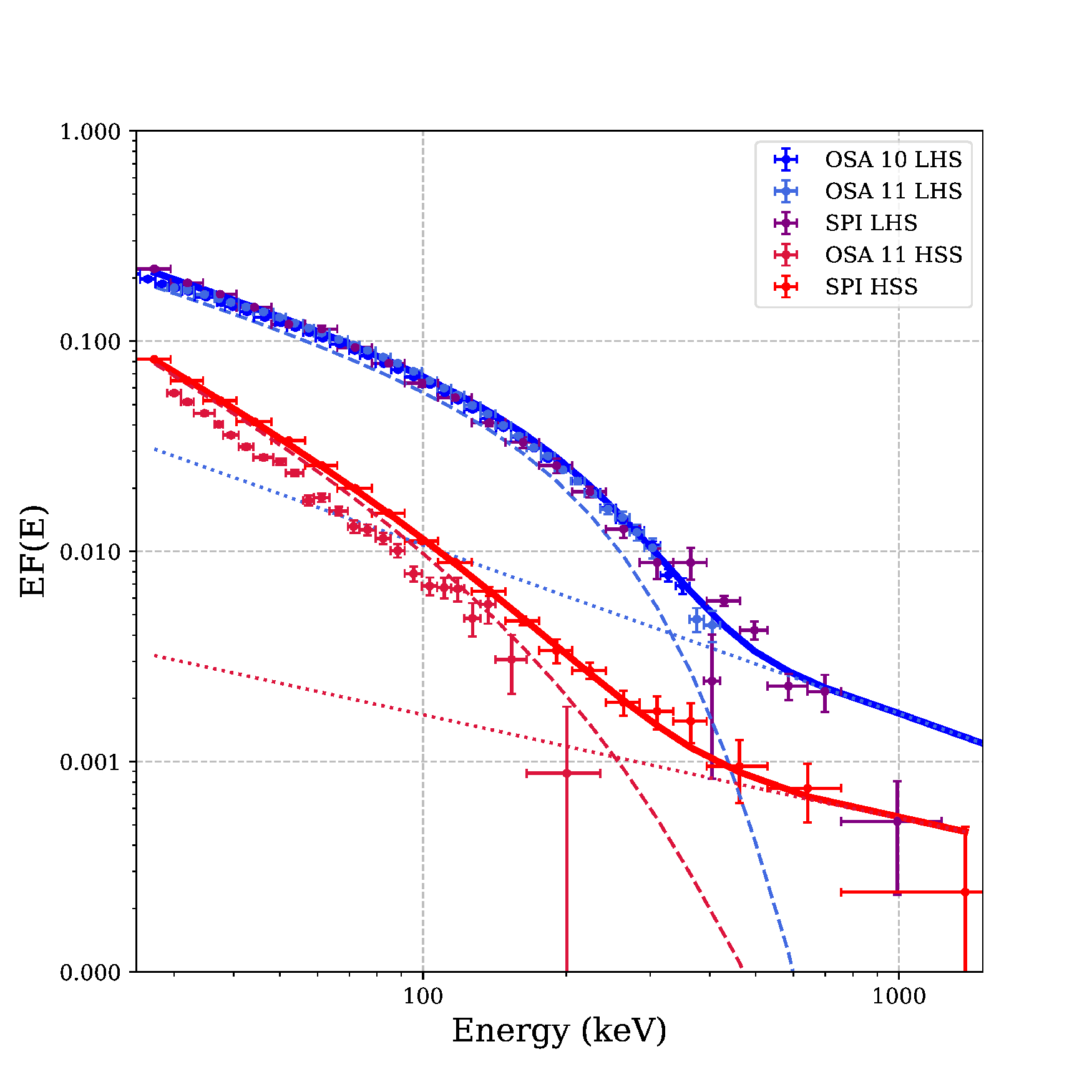}}
\vspace{-0.5cm}
\caption{\footnotesize 
Spectral results. LHS shown in blue and HSS in red.
For both, we show the semi-physical model (\textit{Model 2)} for the HSS) in thick lines. The Comptonisation and power law component are represented in dashed and dotted lines, respectively.
}
\label{spectra}
\end{figure}

\textbf{HSS}. 
Compared to previous studies, the larger amount of data for the HSS allows us to claim a significant excess at high energies. Below\,400 keV, we started first with a simple power-law model  (\texttt{const*powerlaw}) to investigate a possible cutoff, which did not describe the observation well (275.34/52 dof). We thus used two alternative approaches:

\textit{Model 1)} Add reflection, which improved the fit significantly (148.2/51 $\chi^2$/dof). We found $\Gamma$=2.38$\pm$0.02 and a reflection component of $\Omega / 2\pi$=0.81$^{+0.18}_{-0.16}$. We froze the values of $\Gamma$ and $\Omega / 2\pi$ and added the data above 400 keV resulting in 160.8/62 $\chi^2$/dof. Letting $\Gamma$ and/or $\Omega / 2\pi$ free did not improve the fit. 

\textit{Model 2)} Using a cutoff power law instead of a simple power law (115.5/51 $\chi^2$/dof), we found $\Gamma_{\mathrm{cut}}$=2.27$\pm$ 0.03 with an exponential cutoff $E_{\mathrm{cut}}$=231$^{+40}_{-31}$\,keV. When adding data above 400\,keV, we observe significant excess emission above the just described model with 135.7/62 $\chi^2$/dof. Therefore, we added a high-energy power law component:
\begin{itemize}
\item $\Gamma_{\mathrm{cut}}$ and $E_{\mathrm{cut}}$ frozen: 123.2/60 $\chi^2$/dof with $\Gamma_{\mathrm{po}}$=0.094$^{-0.8}_{+0.5}$
\item $\Gamma_{\mathrm{cut}}$ frozen and $E_{\mathrm{cut}}$ free: 120.4/59 $\chi^2$/dof with $\Gamma_{\mathrm{po}}$=0.54$_{-0.8}^{+2}$, $E_{\mathrm{cut}}$= 223$^{+16}_{-38}$ keV
\item $\Gamma_{\mathrm{cut}}$ and $E_{\mathrm{cut}}$ free: 114.7/58 $\chi^2$/dof  with $\Gamma_{\mathrm{po}}$=1.5$_{-0.4}^{+0.9}$, $E_{\mathrm{cut}}$=135$^{+54}_{-42}$ keV
\end{itemize}
Reflection components do not improve these fits. We prefer the \emph{model 2)} with $\Gamma_\mathrm{cut}$ and $E_\mathrm{cut}$ are left free to vary from a statistical point of view.

\section{Semi-physical spectral analysis}
\textbf{LHS}. Starting from the phenomenological model, we replaced \texttt{cutoffpl} by a Comptonisation model (\texttt{comptt}). When freezing $\Gamma_{\mathrm{po}}$ from the phenomenological approach, we found $kT$=55$\pm$2\,keV and $\tau$=0.95$\pm$0.05 (215.9/98 $\chi^2$/dof). The statistics are improved when fitting $\Gamma_{\mathrm{po}}$ freely. We found $\Gamma$=1.8$\pm$0.1,  $kT$=51$\pm$3 keV and $\tau$=1.05$^{+0.1}_{-0.08}$ (205.7/97 $\chi^2$/dof). The values of the Comptonisation parameter $y$ are compatible (0.39$\pm$0.03 and $y$=0.44$\pm$0.06 for $\Gamma$ fixed and free). Adding reflection while freezing $\tau$ improved the fitting (191.3/97 $\chi^2$/dof), we found $\Omega/2\pi$=0.23$_{-0.11}^{+0.13}$, $kT$=55$\pm$3 keV and $\Gamma_\mathrm{po}$=1.7$_{-0.2}^{+0.1}$.

\textbf{HSS}. Starting from \textit{Model 1)}, we froze the reflection component and replaced the power law by the \texttt{comptt} model. We obtained 326.0/59 $\chi^2$/dof and thus let the reflection component free. We found $\Omega/2\pi$=0.13$_{-0.13}^{+0.15}$ with $kT$=246$_{-23}^{+68}$ keV and $\tau$=0.018$_{-0.004}^{+0.03}$ and a better statistic of 133.4/59  $\chi^2$/dof.  Then we froze $\tau$ and added a power law component to model the high-energy excess. This improved the fit (124.7/58  $\chi^2$/dof), and we found $\Gamma$=-0.3$\pm$1.1 with $kT$=246$\pm$7 keV and $\Omega/2\pi$=0.15$^{+0.11}_{-0.10}$.

Starting from \textit{Model 2)}, we replaced the cutoff power law component by a Comptonisation component. By fixing the photon index to the value found in the first approach, we found: 113.7/60 $\chi^2$/dof and $kT$=78$\pm$14 keV, $\tau$=0.21$^{+0.08}_{-0.05}$. Freeing $\Gamma_{\mathrm{po}}$ provides the same spectral parameters.

\section{Polarimetry analysis}
We note that the following analysis is preliminary. We carry out the polarisation analysis using the Compton mode \citep{forot_compton_2007}. We first analysed the full 400--2000\,keV data for LHS and HSS to investigate the global polarisation of the high-energy tail. For the LHS, we detected polarisation with a polarisation angle of 40$\pm$12$^\circ$ and a polarisation fraction of 40$\pm$15\%. Our results are compatible with L11 and R15 but better constrained. The polarisation angle is consistent with the result from SPI, but our polarisation fraction is significantly lower (\cite{jourdain_emission_2012} found PA=40$\pm$3$^\circ$ and PF$>$75\%). For the HSS, no polarisation is detected with an upper limit of 51\%.

Since the high-energy tail is variable (e.g. R15), we divided the whole data set into six different epochs of 932 days each and measured polarisation for each of these six epochs in a state-resolved manner. We found that the first and third epoch, both in LHS, between MJD 52722--53553 and MJD 54575--56439, respectively, are polarized.
These results, however, have to be consolidated before we can conclude to an intrinsic variation of the polarisation. If confirmed, there variations could explain the somewhat lower polarisation fraction we measure.




\section{Conclusion and discussion}
We performed a broad-band spectral study for the LHS and HSS using IBIS and SPI. 
Besides the LHS, we for the first time report a firm detection of a hard tail in the HSS (see \cite{jourdain_integral_2014} for a previous attempt). The systematics result in a large range of $\Gamma_{\mathrm{po}}$= 1.2--1.9 for the LHS and $\Gamma_{\mathrm{po}}$=$-$1.4--2.4 for the HSS. Given the data, the HSS seems to statistically indicate a lower value of $\Gamma_{\mathrm{po}}$. We abstain from providing a firm physical interpretation of these results at this stage of the analysis, especially given the phenomenological description of the hard tail, and systematics arising when averaging over multiple source activity periods.

Using the Compton mode, we measured polarisation of the high-energy tail for both states. The tail is polarised in the LHS with hints of variability of the polarisation fraction over time. Polarisation is not detected in the HSS with 51\% upper limit. 

For the LHS, our results on the polarization properties are compatible with the scenario of synchrotron emission from the jets as proposed by L11, \citep{zdziarski_mev_2012}, R15. For the HSS, assuming that there are still jets but fainter than in the LHS \citep[e.g.][]{drappeau_dark_2016}, the synchrotron tail would be undetectable in the X-ray domain. Alternatively, the hard tail in the HSS could be explained by non-thermal Comptonisation; \cite{romero_coronal_2014} predicted a fraction of polarisation of 54\,\% in state with absence of jets which is roughly compatible with our upper limit of 51\,\%.

The presented results are intermediate and part of a work in progress. We will in particular employ more physical models and consolidate the polarisation analysis. We will also consult radio data as a tool to connect polarisation variability to jet emission.

  
\begin{acknowledgements}
FC, JR, \& PL acknowledge partial funding from the French Space Agency (CNES).
TS is supported by the German Research Society (DFG-Forschungsstipedium SI 2502/1-1). VG is supported through the Margarete von Wrangell fellowship by the ESF
and MWK Baden-W\"urttemberg
\end{acknowledgements}
\vspace{-0.5cm}
\bibliographystyle{aa}
\bibliography{biblio_ok}

\end{document}